\documentclass[]{aastex631}

\newcommand{\eb}{\begin{equation}}
\newcommand{\ee}{\end{equation}}
\newcommand{\masyr}{mas yr$^{-1}$}
\newcommand{\gbp}{$G_{\rm BP}$}
\newcommand{\grp}{$G_{\rm RP}$}
\newcommand{\phe}{\texttt{phot\_bp\_rp\_excess\_factor}}
\newcommand{\baz}{\text{\reflectbox{$\mathsf{z}$}}}
\definecolor{rkka}{RGB}{219,66,32}

\shorttitle{Double and lensed quasars}
\shortauthors{Makarov \& Secrest}

\begin{document}

\title{A catalog of candidate double and lensed quasars from Gaia and WISE data}

\correspondingauthor{Valeri V. Makarov}
\email{valeri.makarov@gmail.com}

\author[0000-0003-2336-7887]{Valeri V. Makarov}
\affiliation{U.S. Naval Observatory, 3450 Massachusetts Ave NW, Washington, DC 20392-5420, USA}

\author[0000-0002-4902-8077]{Nathan J. Secrest}
\affiliation{U.S. Naval Observatory, 3450 Massachusetts Ave NW, Washington, DC 20392-5420, USA}

\begin{abstract}
Making use of strong correlations between closely separated multiple or double sources and photometric and astrometric metadata in Gaia EDR3, we generate a catalog of candidate double and multiply imaged lensed quasars and AGNs, comprising 3140 systems. It includes two partially overlapping parts, a sample of distant (redshifts mostly greater than 1) sources with perturbed data, and systems resolved into separate components by Gaia at separations less than $2\arcsec$. For the first part, which is roughly one third of the published catalog, we synthesized 0.617 million redshifts by multiple machine learning prediction and classification methods, using independent photometric and astrometric data from Gaia EDR3 and WISE with accurate spectroscopic redshifts from SDSS as a training set. Using these synthetic redshifts, we estimate a rate of 4.9\% of interlopers with spectroscopic redshift below 1 in this part of the catalog. Unresolved candidate double and dual AGNs and quasars are selected as sources with marginally high BP/RP excess factor (\phe), which is sensitive to source extent, limiting our search to high-redshift quasars. For the second part of the catalog, additional filters on measured parallax and near-neighbor statistics are applied to diminish the propagation of remaining stellar contaminants. The estimated rate of positives (double or multiple sources) is 98\%, and the estimated rate of dual (physically related quasars) is greater than 54\%. A few dozen serendipitously found objects of interest are discussed in more detail, including known and new lensed images, planetary nebulae and young infrared stars of peculiar morphology, and quasars with catastrophic redshift errors in SDSS.
\end{abstract}

\section{Introduction} \label{section: Introduction}


Dual AGNs and quasars are of tremendous importance to several fields of astrophysics. In extragalactic astronomy, there seems to be a clear evolutionary path for physically interacting binary supermassive black holes (SMBHs) to emerge via the merger of early galaxies or their building blocks \citep{2000MNRAS.311..576K,2011ASL.....4..181C}. They are believed to eventually coalesce, generating the most powerful bursts of gravitational radiation in the universe. The details of this process are far from clear, however. The initial separation, on the scale of $10^3$--$10^4$~pc, and the initial relative velocity are too high for a direct interaction between the BHs. Dynamical friction 
within a gaseous medium is believed to be responsible for the initial hardening, which would bring the BHs within 1~pc or less \citep[e.g.,][]{2013CQGra..30x4008M}. The role of stars or star formation is not well understood, nor the time scales 
or the conditions for both BHs to form independent accretion disks and be active at the same time. It is not known if widely separated double AGNs are much more common at earlier cosmological epochs than in the local universe. The rate of galactic mergers 
appears to peak at redshifts $z\sim 2-3$, up from one-fifth of this peak value around $z\sim 6$ \citep[e.g.,][]{2019A&A...631A..87V}. Mergers at moderate redshifts have also been shown to have a significantly higher rate of AGNs than single galaxies \citep[e.g.,][]{2020A&A...637A..94G}, supporting the evolutionary picture of galaxy mergers triggering BH activity \citep[e.g.,][]{2008ApJS..175..356H}.




Previously published collections of candidate physically double (lensed or dual) quasars explored the data from the SDSS \citep{2006AJ....131.1934I,2012AJ....143..119I} and are limited to $\sim 10^2$ objects, with even fewer confirmed duals at close separations \citep[see Figure~8 in][]{2017ApJ...848..126S}. A promising method that leverages the massive statistical power of SDSS spectroscopy, selection on double-peaked [\ion{O}{3}] emission, is nonetheless very inefficient, with most objects being revealed as single AGNs in follow-up studies \citep[e.g.,][]{2010ApJ...716..131R, 2015ApJ...813..103M, 2020ApJ...892...29F}. Aside from successful X-ray campaigns in nearby AGNs \citep{2012ApJ...746L..22K}, high angular resolution radio studies with the VLBA and VLA have generally been required to confirm the presence of dual compact, flat-spectrum radio AGNs \citep[e.g.,][]{2006ApJ...646...49R, 2011ApJ...740L..44F, 2017NatAs...1..727K}, although pointed observations are observationally inefficient and the majority of AGNs are not expected to be radio-bright. Nonetheless, research on dual quasars and AGNs is gaining momentum because of their importance as the progenitors of powerful gravitational wave events that serve as ``standard sirens'', the gravitational equivalent to standard candles such as Type~Ia SNe \citep[e.g.,][]{2005ApJ...629...15H, 2010RvMP...82.3069C}.
As quasars generally reside in galaxies, they have peculiar motions that can be several hundreds of kilometers per second relative to the Hubble flow. The resulting astrometric proper motions should be negligible for most of them because of the great distances separating the sources from the observer, and indeed this is a major reason for their use in creating the International Celestial Reference Frame \citep[ICRF;][]{2020A&A...644A.159C}, the physical realization of the ICRS coordinate system. For a standard flat $\Lambda$CDM cosmology with $H_0=70$~km~s$^{-1}$~Mpc$^{-1}$ and $\Omega_\mathrm{m}=0.3$, one milliarcsecond subtends 8~pc at a typical quasar redshift of $z=1$, so a quasar with a peculiar velocity of 100~km~s$^{-1}$ will have an intrinsic proper motion of $\sim0.01$~$\mu$as~yr$^{-1}$, three orders of magnitude below even the most precise proper motions available from Gaia.

Multiplicity is one of the important factors that can perturb precision astrometry of quasars \citep{2012MmSAI..83..952M}. The closer AGNs at redshifts $z\lesssim 0.5$ usually have resolved host galaxies associated with them, which are in general asymmetric to some degree. Quasars found in double systems at large physical separations \citep{2006AJ....131.1934I,2009ApJ...705L..76W,2012ApJ...753...42C} presumably originate from mergers of galaxies with individual central AGNs. The typical scale of projected separations is $\sim1$~kpc, with a wide distribution of a few orders of magnitude. Cosmological simulations predict that the incidence of dual AGNs at moderate redshift is of order one to a few percent \citep[for a review, see][]{2019NewAR..8601525D}. At a much lower rate, quasars can be gravitationally lensed by foreground galaxies with doubly or multiple-imaged configurations \citep{2019A&A...622A.165D}. The lensed images of currently known systems are mostly packed within a few arcseconds. A yet unknown fraction of quasars at high redshifts are multiply-imaged systems lensed by more distant and less massive lenses that subtend $1\arcsec$ or less. These features can likely produce astrometric position offsets. Bogus proper motions can be measured in some cases for unresolved lensed images of variable quasars because of the time lag of the light curve due to light travel time and gravitational delay effects \citep{2012A&A...544A..51K} and resulting photocenter shifts. Quasars that exhibit apparent proper motions are rare objects, and have only recently begun to be studied \citep{2022A&A...660A..16S, 2022ApJ...933...28M}.

The main objective of this paper is select candidate dual AGNs (CDAGNs) from mid-IR AGNs (hereafter ``MIRAGNs'') in the catalog of \citet{2015ApJS..221...12S}, cross-matched with Gaia~EDR3. Using a mid-IR-selected initial sample favors obscured quasars at higher redshifts, and therefore results in a helpful bias toward galactic mergers \citep{2020A&A...637A..94G}. Smaller test samples of quasars with spectroscopically determined redshifts from the Sloan Digital Sky Survey \citep[SDSS][]{2020ApJS..250....8L} and available Pan-STARRS images \citep{2010SPIE.7733E..0EK} are used in this paper as independent verification of our detection criteria. We also compute the near-neighbor distance statistics of resolved companions within $11\arcsec$ for a large sample of 0.632 million quasars present in Gaia~EDR3 to confirm the presence of real binary AGNs. Our chosen method of selecting unresolved CDAGNs is vulnerable to perturbations in the Gaia data caused by extended structures around central sources at low to moderate redshifts. Our search is therefore limited to $z>1$ objects. Since accurate spectroscopic redshifts are only available for $\sim 23\%$ of the MIRAGN sample that is located within the SDSS footprint, we develop and employ a number of Machine Learning (ML) techniques to estimate redshifts from the more widely available data from Gaia EDR3 and WISE. In both the prediction and classification regimes, the ML-generated data are trained on and compared with the SDSS redshifts.

\section{Methodology} \label{section: Methodology}
\subsection{Initial quasar sample}
We cross match the catalog of 1.4~million MIRAGNs from \citet{2015ApJS..221...12S} to the Gaia~EDR3 catalog, using a match tolerance of $0\farcs5$ for reliability. This tight search radius causes a loss of some genuine matches due to the limited astrometric precision of WISE, but it helps to reduce the rate of source confusion in resolved pairs. This produced 621,946 matches, 551,482 of which have valid proper motion and parallax measurements. For verification and machine learning training purposes, we also match the Gaia counterpart coordinates on to the SDSS specObj-dr16.fits table,\footnote{\url{https://www.sdss.org/dr16/spectro/spectro_access/}} allowing only spectra with \texttt{ZWARNING==0} or \texttt{ZWARNING==4}, the latter of which can happen for spectra with broad lines.\footnote{\url{https://www.sdss.org/dr16/spectro/caveats/\#zstatus}} To ensure spectroscopic fiber coverage of the Gaia counterpart for this auxiliary test sub-sample, we allow BOSS spectra within $1\arcsec$ and SDSS spectra within $1\farcs5$, resulting in 126,343 objects. The high reliability of the initial selection---prior to matching on the SDSS spectra table---is illustrated by the extremely low rate of stellar contaminants with only 72 sources having negative SDSS redshifts. The total number of objects spectroscopically classified as ``STAR" is 135 ($0.1\%$).

\subsection{Redshift training parameters}
Because the SDSS footprint is about one quarter of the full sky, accurate spectroscopic redshifts are available for just 23\% of the initial sample. To increase the output of our study, we generate synthetic redshifts using statistically correlated data from WISE and Gaia as training parameters. Ideal dependencies would include monotonic functions of redshift with a negligibly small dispersion. Needless to say, there are no astrometric or photometric parameters that possess such properties because of the multitude of complex physical phenomena that contribute to the available observables. We reviewed dozens of parameters available in the WISE and Gaia catalogs looking for any correlations with redshifts in the sample of 0.126 million sources and selected the few most promising ones. For the prediction machine learning (ML) training set, we settled on just four purely photometric parameters.

\begin{figure*}
    \includegraphics[width=0.47 \textwidth]{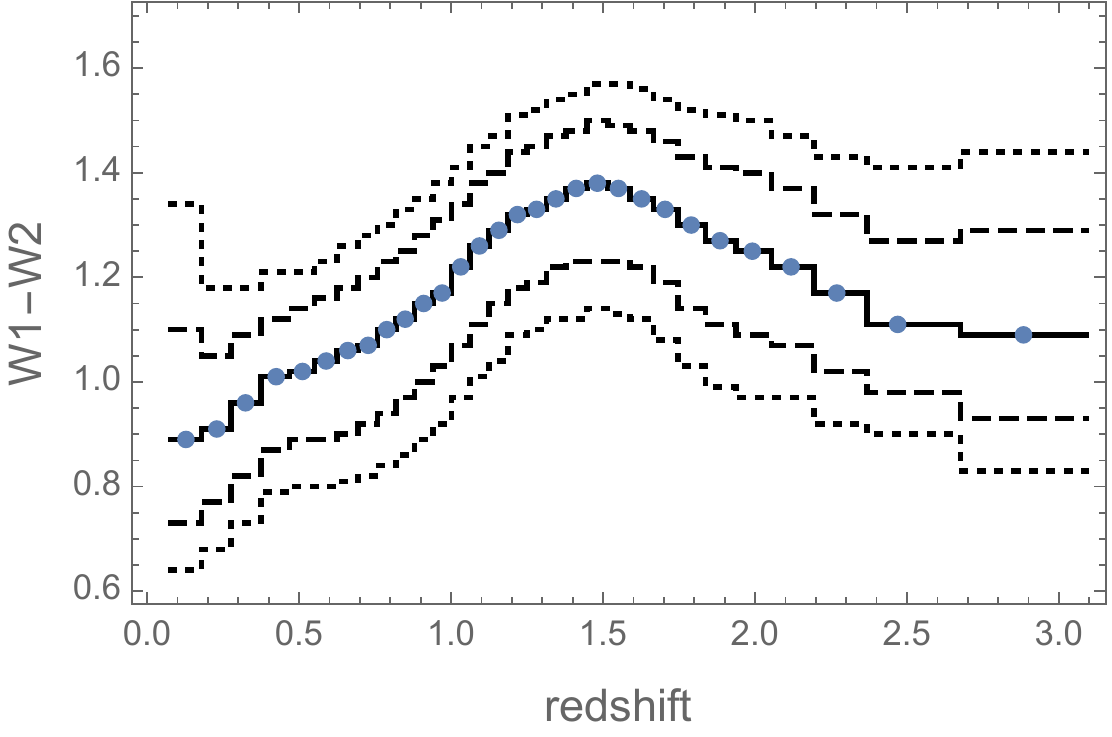}
    \includegraphics[width=0.47 \textwidth]{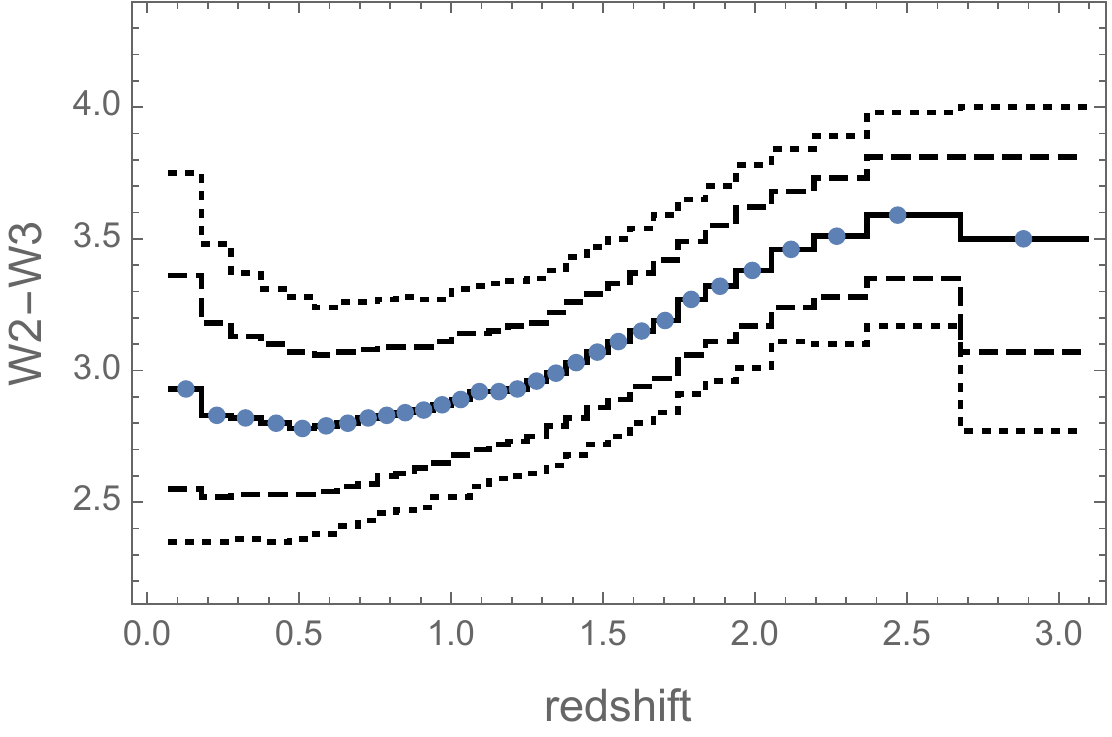}
    \caption{Statistical dependence of WISE mid-IR colors on spectroscopic redshift.
    The broken solid line with dots shows the median values in 30 equal bins of sorted redshifts. Each dot indicates the median value of redshift for the corresponding bin, which is not centered within the bin. The dashed lines show the 0.16--0.84 intervals of colors in the same binned sub-samples. The dotted lines represent the 0.05--0.95 interval. Left: $W1-W2$ magnitudes; Right: $W2-W3$ magnitudes.}
    \label{w.fig}
\end{figure*}

Figure~\ref{w.fig} shows the dependencies of two photometric training parameters from the WISE mission data on SDSS-determined redshifts. These mid-IR magnitudes have been used to generate the input catalog of MIRAGN \citep{2015ApJS..221...12S}, which serves as our starting sample. The overall distribution of mid-IR colors is quite red by selection. The plots are generated by sorting the sample of 0.126 million sources by spectroscopic redshift, dividing it into 30 non-overlapping bins of equal size, and computing for each bin the median redshift and the median color (shown with a solid broken line with data points) as well as the $\{0.16, 0.84\}$ quantiles (shown with dashed lines) and $\{0.05, 0.95\}$ quantiles (dotted lines).
The former pair of quantiles is the robust statistics analog of the $\pm 1\sigma$ interval for a Gaussian distribution of probabilities. 
In both cases, the scatter of color around the median value for a given redshift is larger than $\gtrsim 0.1$ mag, which limits the performance of the ML prediction, deteriorating most notably for the nearest sources. While the $W2-W3$ color displays a nearly monotonic dependence at $z>1$, the $W1-W2$ color has a distinct turnover, peaking at $z=1.5$. Unfortunately, the gradients are rather small, preventing a good performance of the ML prediction on just these two parameters.

\begin{figure*}
    \includegraphics[width=0.47 \textwidth]{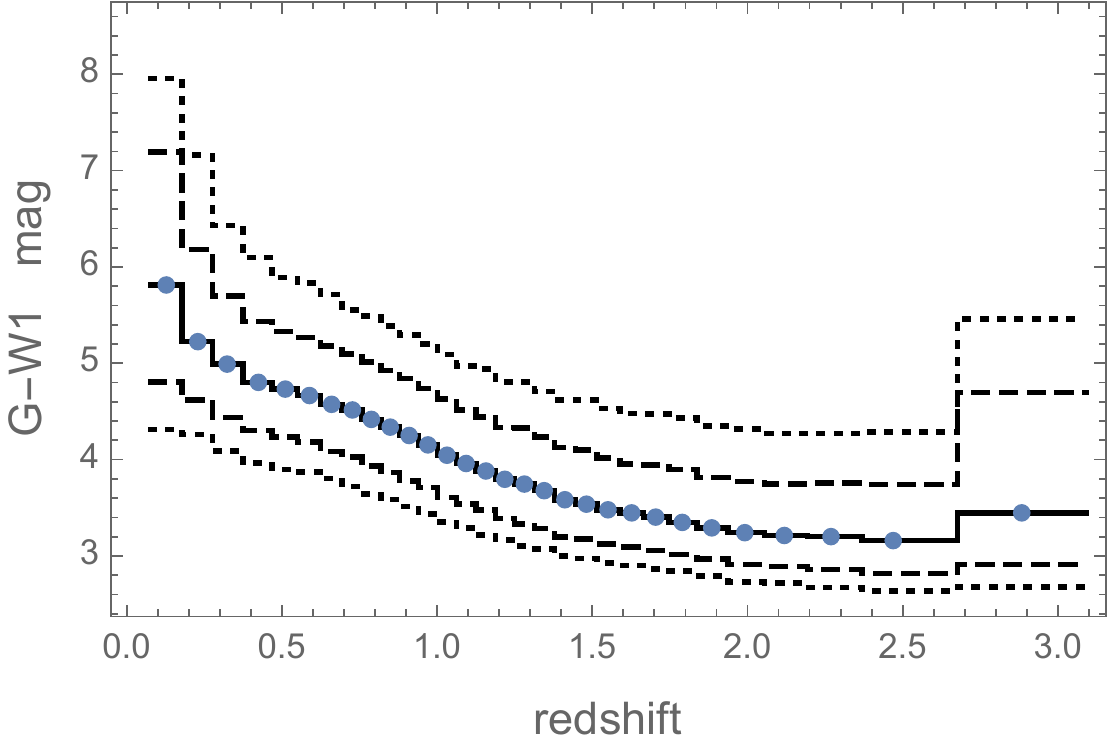}
    \includegraphics[width=0.47 \textwidth]{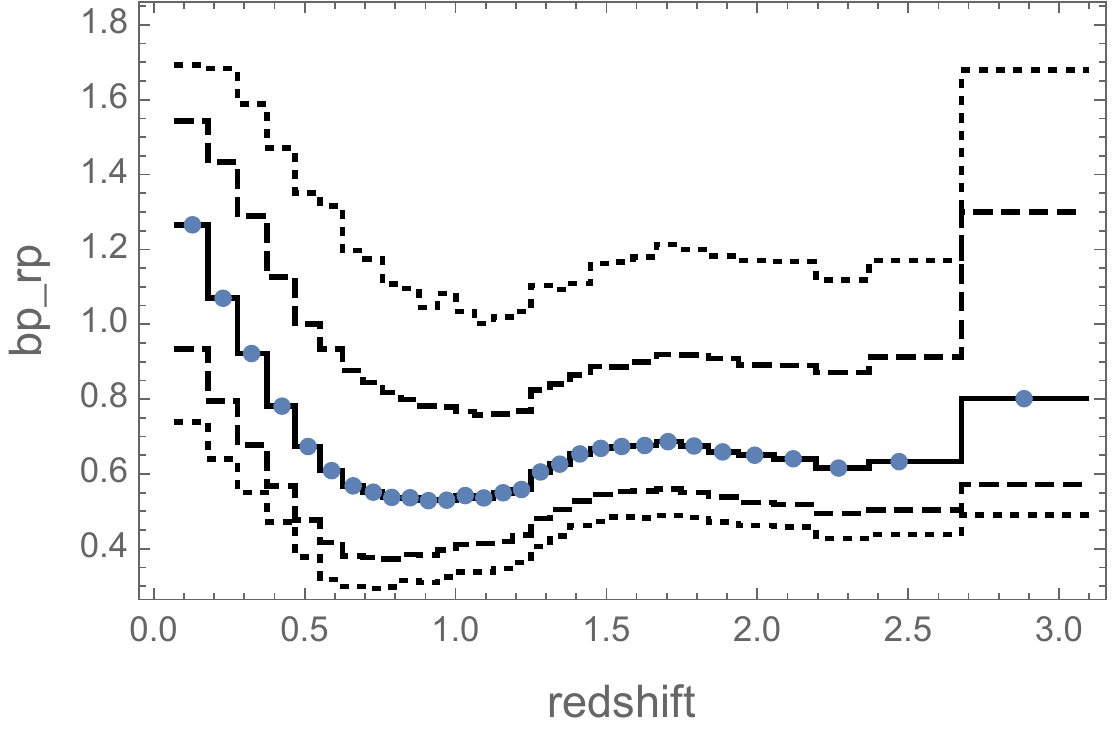}
    \caption{Statistical dependence of optical/mid-IR colors on spectroscopic redshift.
    The broken solid line with dots shows the median values in 30 equal bins of sorted redshifts. The dashed lines shows the 0.158655 and  0.841345 quantiles of colors in the same binned sub-samples. The dotted lines represent the 0.05 and 0.95 quantiles. Left: $G-W1$ magnitudes; Right: $G_{\rm BP}-G_{\rm RP}$ magnitudes.}
    \label{g.fig}
\end{figure*}

A more pronounced gradient is the $G-W1$ color with redshift, albeit with larger scatter, shown in Figure~\ref{g.fig}, left. Nearby AGNs are distinctly redder then their more distant counterparts. The median $G$ magnitude,
on the other hand, is quite flat across the range of redshift. We interpret this drop in $G-W1$ by approximately 2.5 magnitudes as an observational selection effect. A large fraction of AGNs is heavily obscured at all redshifts, and due to relatively shallow optical magnitude limit of Gaia ($G\lesssim20$), become too faint or completely invisible with increasing distance. The fourth parameter used for the ML redshift prediction is the Gaia EDR3 optical color $G_{\rm BP}-G_{\rm RP}$, given as the {\texttt bp\_rp} column in the Gaia catalog (Figure~\ref{g.fig}, right). It is probably the weakest discriminator because of the fairly flat and non-monotonic dependence at $z>0.5$. We include it hoping to get a better performance for objects at the smallest redshifts, which exhibit the highest rate of large errors, as explained below.

The main source of perturbation of Gaia astrometric and photometric measurements is not double or multiple morphology (which is a relatively rare occurrence) but the extended character of host galaxies at small redshifts. All sources in Gaia EDR3 are indiscriminately processed as point-like stellar images using a unified set of Line Spread Functions (LSF). The drastic deterioration caused by the extended structures at $z<0.5$ is illustrated by the analysis of the photometric excess parameter \phe\ in Makarov \& Secrest 2022. Because of the different methods used in EDR3 to estimate the broad band magnitudes $G$ and the narrower \gbp\ and \grp\ magnitudes, the ratio of the corresponding fluxes may deviate upward significantly from the most common value slightly above 1 \citep{Riello21}. While the majority of more distant and luminous quasars tightly group around a well-defined lower envelope at \phe$\simeq 1.1$, the closer AGNs show a dramatic increase both in the median and dispersion values. The more luminous AGNs detectable at high redshift have a higher ratio of AGN$/$galaxy emission, so that the contribution of underlying host structures quickly tapers off with $z$. On the other hand, as was demonstrated by Makarov \& Secrest 2022, \phe\ is the most reliable indicator of resolved or unresolved multiplicity of quasar images at $z>1$. As the goal of this investigation is to remove sources at $z<1$ as cleanly as possible without removing double or multiple sources at higher redshift, we do not use \phe\ as a parameter in our ML training set.

\subsection{Machine learning prediction of $z$}
\label{mlpre.sec}
Having selected the four photometric training parameters, we generate a sample of 0.126 million sources with accurate SDSS spectroscopic redshifts, which is a 20.5\% subset of our working sample counting 0.617 million objects. The subset of sources with spectroscopically determined redshifts is randomly divided into two equally sized parts. The first is the training set, on which the ML classifier is tuned to predict the observed $z$ from the data vector $\{W1-W2,\,W2-W3,\,G-W1,\,$\gbp-\grp\}. The second is the test set for assessing the performance of the different ML methods discussed below. This assessment is based on two statistical metrics of interest. The first is the robust standard deviation proxy 1.5 times the median absolute deviation (MAD) of the measured versus predicted redshifts $z_{\rm obs}-z_{\rm pre}$. This allows for determination of the dispersion of $z_{\rm obs}-z_{\rm pre}$ independent of the presence of statistical outliers. The rate of interlopers is the second statistical metric of interest, which we define as objects with $z_{\rm pre}>1$ and $z_{\rm obs}<1$, and separately quantify the rate of more dangerous contaminants with $z_{\rm pre}>1$ and $z_{\rm obs}<0.5$. Note that while the training is performed on a set of 0.063 million randomly selected sources with observed redshifts, the trained prediction or classification is applied to the entire working sample of 0.617 million objects, including the training and test subsets.

\begin{figure*}
    \includegraphics[width=0.48 \textwidth]{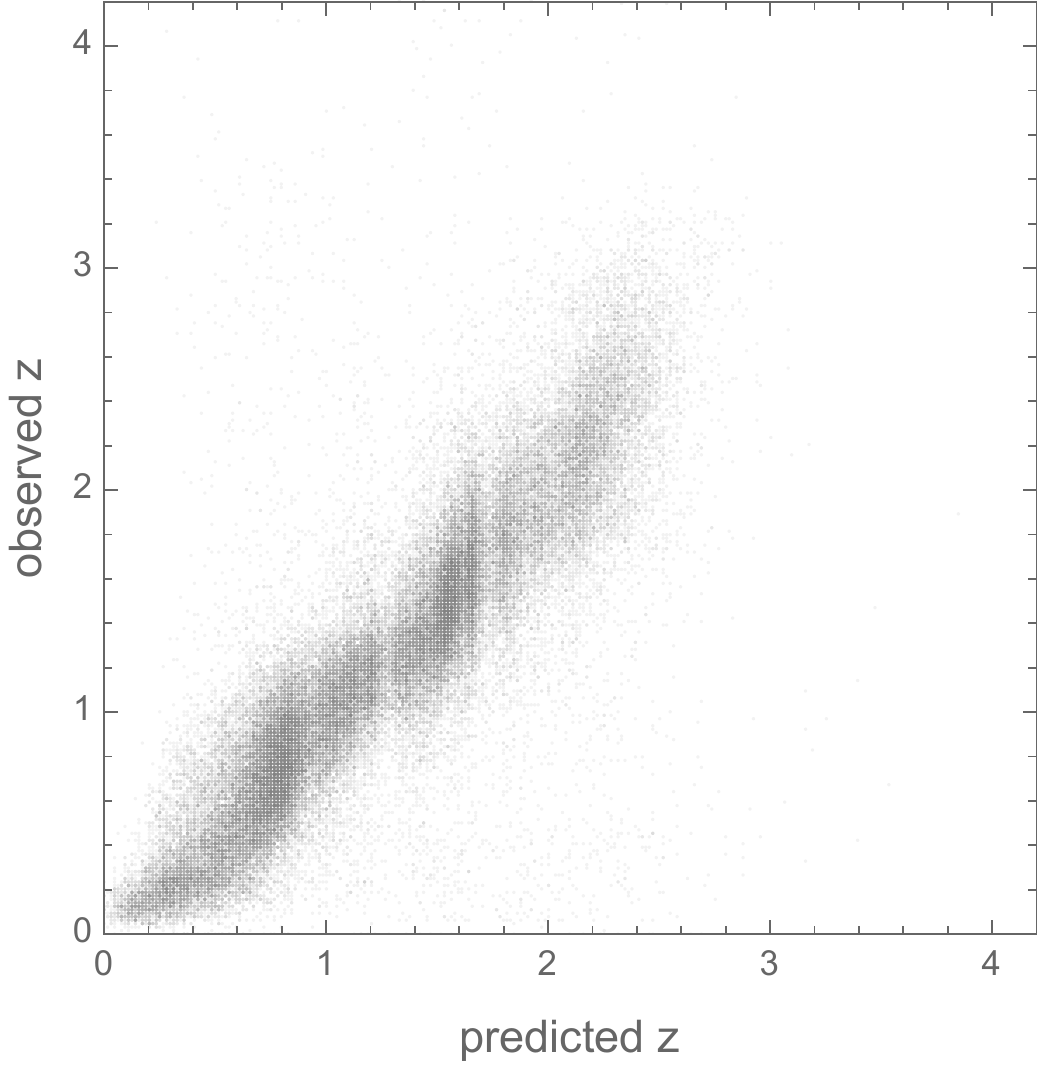}
    \caption{Machine Learning predicted redshifts (with the Neural Network method) versus SDSS spectroscopic redshifts for a test sample of 0.063 million quasars.}
    \label{ml.fig}
\end{figure*}

We experimented with ML classifiers using six different prediction methods: Linear Regression, Gradient Boosted Trees, Decision Trees, Nearest Neighbors, Random Forest, and Neural Network. Multiple ML training and prediction runs were implemented with randomly selected training sets and different methods. The performance of each run was evaluated using the two metrics. The best results were obtained with the Nearest Neighbors and Neural Network, and the latter was accepted for processing the entire sample. The best trials yielded a rate of interlopers of 5.64\% and a rate of contaminants of 0.93\%. Figure~\ref{ml.fig} depicts the distribution of $z_{\rm pre}$ versus $z_{\rm obs}$ for the test set. The robust standard deviation of the differences is 0.235. With 63,000 objects in the test set, we note a statistically significant bias of the predicted values with a median $z_{\rm obs}-z_{\rm pre}$ of $-0.036$. If such objects propagate into our final selection at a significant rate, they can dominate the final selection of candidate double quasars.

Since we are only interested in sources with redshifts greater than 1, we also employed a range of ML classification methods, which provide a binary classification (yes or no). The choice of available techniques is wider, and the logic is somewhat different. Therefore, classification and prediction outcomes are not identical for the same data sample and training set. This redundancy helps us to achieve a slightly higher reliability of the selected sample.

\subsection{Machine learning classification of $z$}
\begin{figure*}
    \includegraphics[width=0.48 \textwidth]{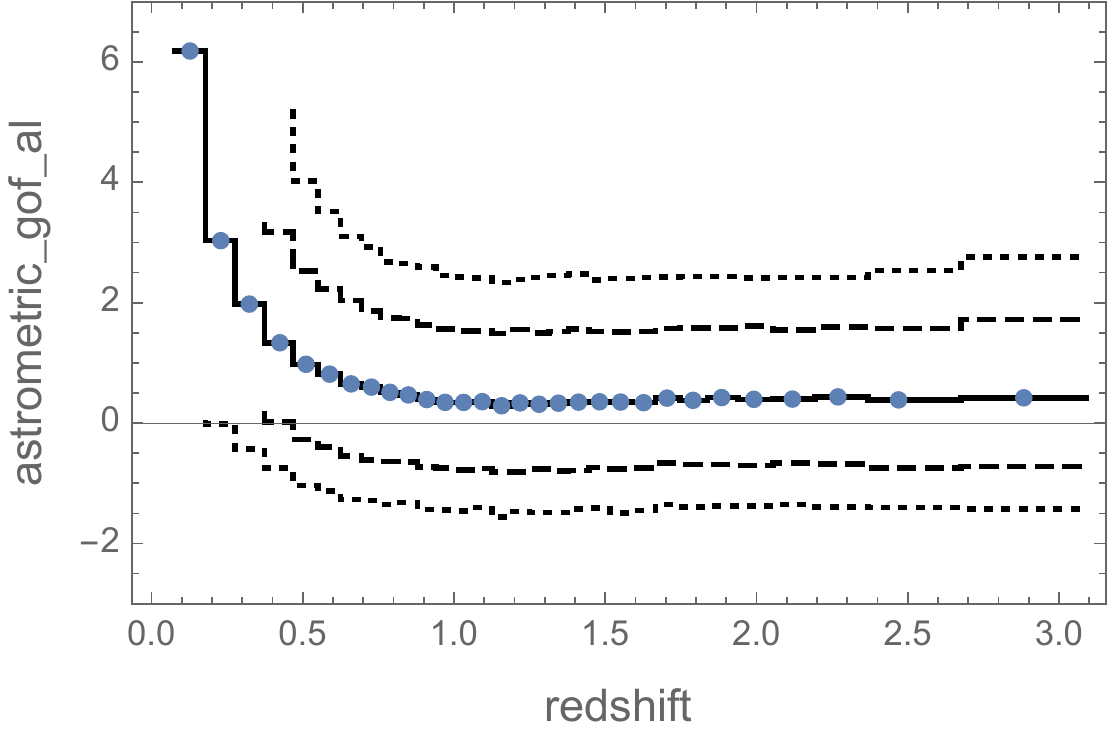}
    \caption{Statistical dependence of Gaia~EDR3 parameter \texttt{ astrometric\_gof\_al} on spectroscopic redshift.
    The broken solid line with dots shows the median values in 30 equal bins of sorted redshifts. The dashed lines shows the 0.16 and 0.84  quantiles of colors in the same binned sub-samples. The dotted lines represent the 0.05 and 0.95 quantiles.}
    \label{gof.fig}
\end{figure*}

A wider range of ML classification methods was applied to the training set in order to separate the sources into categories $\baz_{\rm cla}=0$ and $\baz_{\rm cla}=1$, which are defined as $z<1$ and $z>1$, respectively. To improve the performance of this classification, an additional astrometric control from the Gaia EDR3 catalog was used. Figure~\ref{gof.fig} shows the statistical dependence of the \texttt{astrometric\_gof\_al} astrometric goodness-of-fit on redshift for the training set. This Gaia parameter indicates the degree of astrometric perturbation in the post-fit residuals with respect to the standard 5-parameter model (position components, parallax, proper motion components). An ideal fit to the data with a dispersion of residuals in agreement with the formal errors would yield a goodness-of-fit value equal to zero. The median dependence is quite flat for $z>1$ but progressively elevated at lower redshift, indicating increasingly more unexplained astrometric variance. A dramatic excess is evident for the nearest AGNs with even the lower 0.05 quantile taking positive values for $z<0.3$. The \texttt{astrometric\_gof\_al} parameter thus provides additional leverage on the nearest sources, which are the most dangerous for this study. The presence of bright extended structures accounts for this behaviour, which makes Gaia astrometric measurements extremely noisy \citep{2017ApJ...835L..30M}. This study also benefits from the apparent lack of correlation between this metadata type and the crucial \phe. The two parameters come from different part of the Gaia pipeline. While \phe\ is sensitive to the registered image structure (multiplicity, extended component), the \texttt{astrometric\_gof\_al} mostly reflects how well the higher-level astrometric measurements comply with the adopted astrometric model.

The Gradient Boosted Trees ML method yielded slightly better classification results than its close competitors Neural Network and Nearest Neighbors. The rate of interlopers (sources with $\baz_{\rm cla}=1$ and and $z_{\rm obs}<1$) came up to 7.60\%, while the loss rate of objects with $z_{\rm obs}>1$ in $\baz_{\rm cla}=0$ amounted to 9.56\%. Although the performance numbers are lower than the ML prediction output (Section \ref{mlpre.sec}), we use these two conceptually independent ML methods to achieve higher reliability of the final catalog. The ML process is intrinsically stochastic, and the borderline objects with redshifts around 1 can obtain different results because of the dispersion of classifier values at a given redshift. Our strategy is to err on the safe side and reject as many such ambiguous cases as possible, at a cost to completeness of our selection.

\subsection{Generating a large sample of high-redshift quasars} \label{large.sec}

Both ML prediction and classification algorithms were applied to a sample of $617,093$ MIRAGN/EDR3 objects that have all five of the training parameters considered in this work (the four photometric parameters, plus \texttt{astrometric\_gof\_al}). 
For each object, two values are obtained, the predicted redshift $z_{\rm pre}$ and the classification flag $\baz_{\rm cla}$. Since the prediction and classification runs are based on different ML methods, use different training sets, and non-identical classifiers, the results are not entirely fully consistent for individual objects, in that the two parameters may be contradictory. In fact, consistent results were obtained for $590,147$ sources, including $303,806$ sources with both indicators placing the redshift above 1 and $286,341$ sources with
both indicators signalling a redshift below 1. Aiming at the highest reliability of the final selection, we only accept the sources in the first category for the catalog. Thus, the general sample is split into 4 non-overlapping subsets, of which we are using in the following processing only two subsets. The first one includes
objects with $z_{\rm pre}>1$ and $\baz_{\rm cla}=1$, hereafter called the Z11 sample. The other part used only for verification purposes includes objects with both $z_{\rm pre}<1$ and $\baz_{\rm cla}=0$, hereafter called the Z00 sample.
Using again the subset of sources with SDSS spectroscopic determinations, a 6.5\% rate of interlopers and a 0.8\% rate of contaminants is estimated for the Z11 sample (see Section~\ref{mlpre.sec}).
At this point, the quality of our Z11 selection is improved by removing 4699 sources with observed redshifts below 1 (i.e., known interlopers), and by adding 8122 objects with observed redshifts above 1 that were missed by our
ML methods (i.e., known losses). The estimated rate of remaining interlopers in the resulting Z11 sample of $307,239$ sources is 4.9\%. A similar cleaning procedure is applied to the Z00 sample.

\begin{figure*}
    \includegraphics[width=0.47 \textwidth]{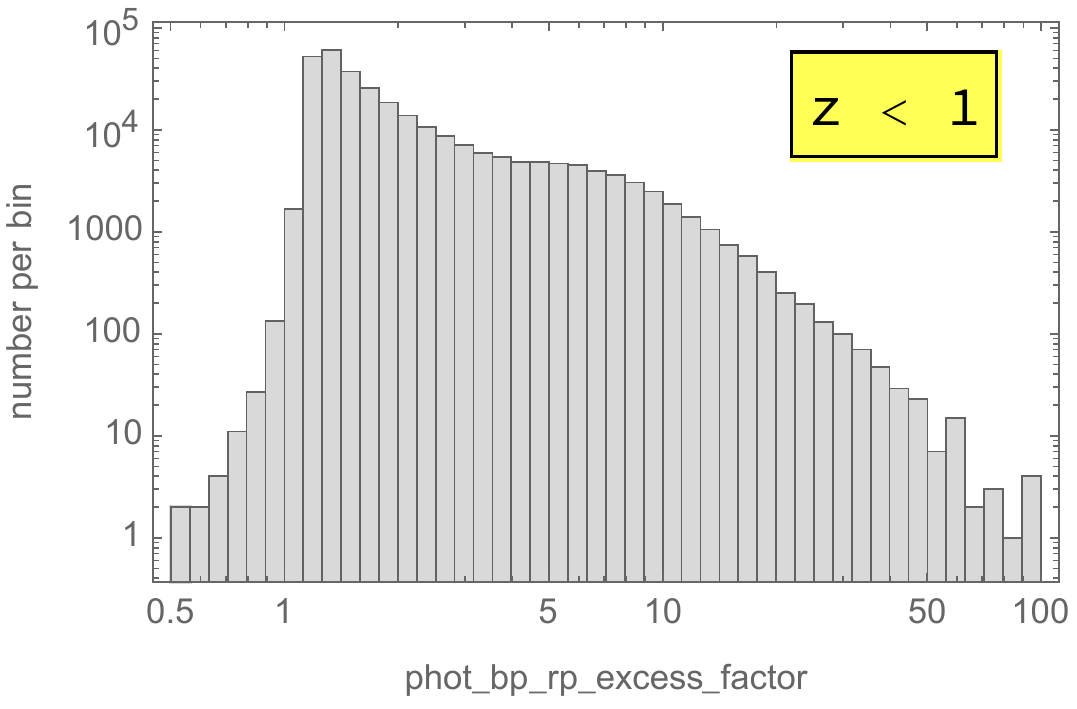}
    \includegraphics[width=0.47 \textwidth]{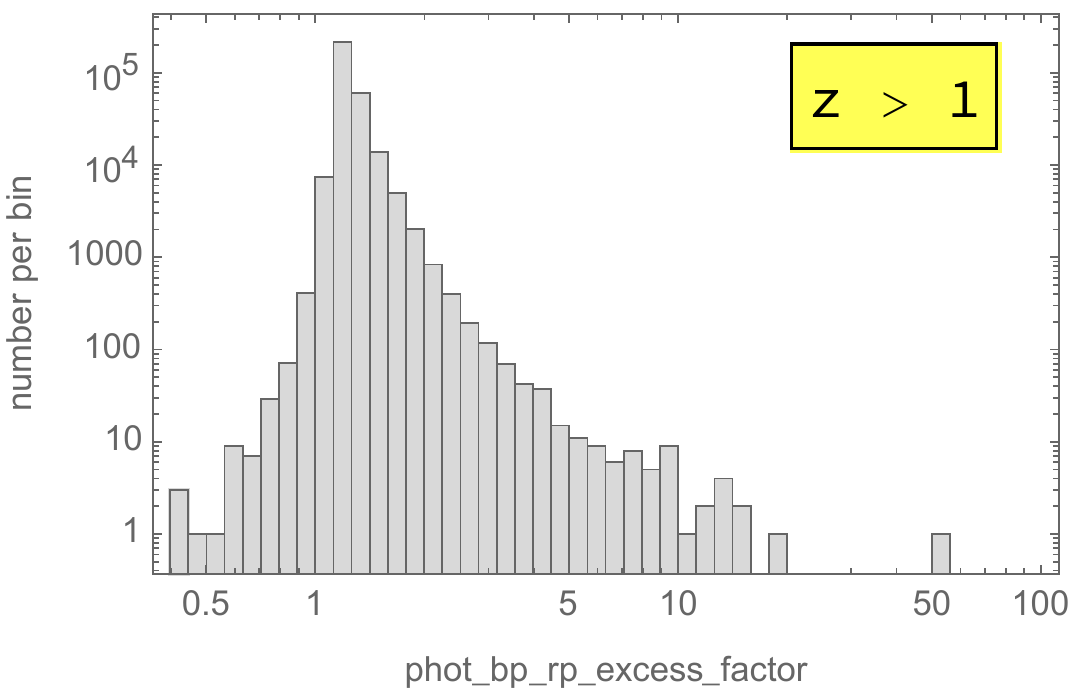}
    \caption{Histograms of the \phe\ parameter for the Z00 (left) and Z11 (right) samples. Note the logarithmic scale of both axes. }
    \label{phe.fig}
\end{figure*}

Figure~\ref{phe.fig} displays the distributions of the \phe\ parameter, which has a crucial role in this study, separately for the Z00 (left plot) and Z11 (right plot) samples. 
We observe significant differences in the width and shape of these histograms. Even though the most frequent values are between 1 and 2 for the nearer sources, the tail of high \phe\ values is much larger than for the more distant quasars, including clear evidence for a secondary population peaked at $\sim5$, absent in the high-$z$ objects.
This confirms the effectiveness of our ML-based selection and filtering of nearby sources with perturbing extended structures.

\section{Selection of candidate dual AGNs by photometric excess}
Makarov \& Secrest 2022 provided evidence that elevated values of \phe\ at $z>1$ are often caused by double or multiple sources in MIRAGN, including optical pairs with Galactic stars, physical dual quasars, and multiply imaged gravitational lenses. We use the Z11 sample of sources with mostly ML-synthesized redshifts and use \phe\ to select candidate double objects that are brighter than $G=20$ mag. This upper limit on the optical magnitude is meant to avoid bogus positives among faint sources where both the photometric ML controls and \phe\ are perturbed  \citep{2021A&A...649A...5F}. 

The lack of bimodality in distribution of \phe\ for high-$z$ quasars in Figure~\ref{phe.fig}, right, precludes the use of a simple threshold criterion. The small excess at very high values, above $\sim5$ is likely to be caused by a small fraction of nearby galaxies and stars with peculiar photometric properties that managed to propagate into the Z11 sample (cf. Appendix A). We therefore performed a series of visual tests by sorting the entire Z11 sample by \phe\ and investigating the rate of Gaia-resolved sources and the appearance of objects in the Pan-STARRS mosaic images at various levels. In this manner, we settled on an empirical sample quantile of 0.995 (leaving only 0.5\%  of the sample with most perturbed values), which corresponds to a threshold value of \phe\ $=1.545$. This is a conservative cut, because tight resolved sources continue to appear at even smaller parameter values, including candidate gravitational lenses listed by \citet{2019A&A...622A.165D}.

The resulting sample, hereafter called Sample~1, counts 1023 sources. We investigated a few possible options to improve its reliability using additional parameters. These include the {\tt phot\_bp\_n\_blended\_transits} and {\tt phot\_rp\_n\_blended\_transits} values, as well as their sum, which may be correlated with the presence of close companions.\footnote{\url{https://gea.esac.esa.int/archive/documentation/GDR3/Gaia_archive/chap_datamodel/sec_dm_main_source_catalogue/ssec_dm_gaia_source.html}} The distributions are very broad without obvious structure. Visual inspection of sources that have the sum of the two parameters below 5 (459 in total) helped to detect a few objects with grossly incorrect SDSS spectroscopic redshifts (Appendix A) without revealing any obvious relation to multiplicity of these sources. Objects with very red optical colors $G_{\rm BP}-G_{\rm RP}>2.6$ that are atypical for quasars at high redshifts were also reviewed. They could be the remaining stellar interlopers mimicking quasars in the mid-IR region of spectra. Indeed, four infrared Galactic objects were identified out of a total of 20, as discussed in Section \ref{stars.sec}. No additional filtering has been applied because of the low rate of identifiable false positives.

\section{Selecting candidate dual AGNs resolved in Gaia EDR3} 
\label{reso.sec}
The second part of CDAGN (hereafter Sample 2) comprises quasars and AGNs that have been directly resolved in Gaia EDR3 irrespective of their redshift. It is straightforward to select all MIRAGN sources with near neighbors. The main problem is how to avoid stellar companions overwhelming this selection given that the majority of optical pairs include stars. 
As discussed in (Makarov \& Secrest 2022), the number of targets without companions within $11\arcsec$ can be used to estimate the effective number density of sources in Gaia EDR3. If $N$ sources are randomly positioned on the celestial globe, the probability of not having at least one companion within an angular radius $r$ (in radians) is 
\eb
P_{\rm empty}(r)=\left( 1-\frac{\pi r^2}{4\pi}\right)^N
\label{pempty.eq}
\ee
This probability is accurately estimated as the ratio of cross-matched MIRAGN objects without any companions to the total number of targets. For $r=11\arcsec$, we obtain $N=3.246\times 10^8$. We can now compute the expected probability of having a random companion within $r=2\arcsec$,
$1-P_{\rm empty}(r=2\arcsec)=0.0076$. We note that such random companions, i.e., optical pairs, would be mainly Galactic stars, because stars dominate in the Gaia catalogs. However, the estimated rate of stellar contaminants within $2\arcsec$ may be biased upward with respect to the modal value by the strongly asymmetric distribution of the rate on the celestial sphere. Indeed, while 79\% of quasars do not have any companions within $r=11\arcsec$, 0.0011\% of the sample have 16 neighbors within the same area. A small fraction of sources in crowded areas (closer to the Galactic plane, projected against globular clusters, etc.) is responsible for a large fraction of chance alignments. We devised a selection approach leading to an inevitable loss of genuine double quasars but producing fewer optical pair contaminants. 

A total of $113,241$ MIRAGN objects have at least one companion within $11\arcsec$, with $98,432$ having only one or two neighbors within this radius. The remaining sources are mostly in areas with high local number density, and their multiple neighbors are likely stars. To eliminate such crowded areas, we filter only sources with neighbors within $5\arcsec$, but without any neighbors between $r=5\arcsec$ and $11\arcsec$. This filtering should remove $1-5^2/11^2\simeq 81$\% of chance alignments in crowded areas leaving $12,049$ candidate sources. For this search of dual AGNs, we are only interested in the tighter separations. Gaia EDR3 resolving capabilities become impacted at separations below $2\arcsec$ \citep{2021A&A...649A...5F}, which is our chosen upper limit for inclusion in CDAGN. The number of thus selected sources is 2363.

\begin{figure*}
    \includegraphics[width=0.47 \textwidth]{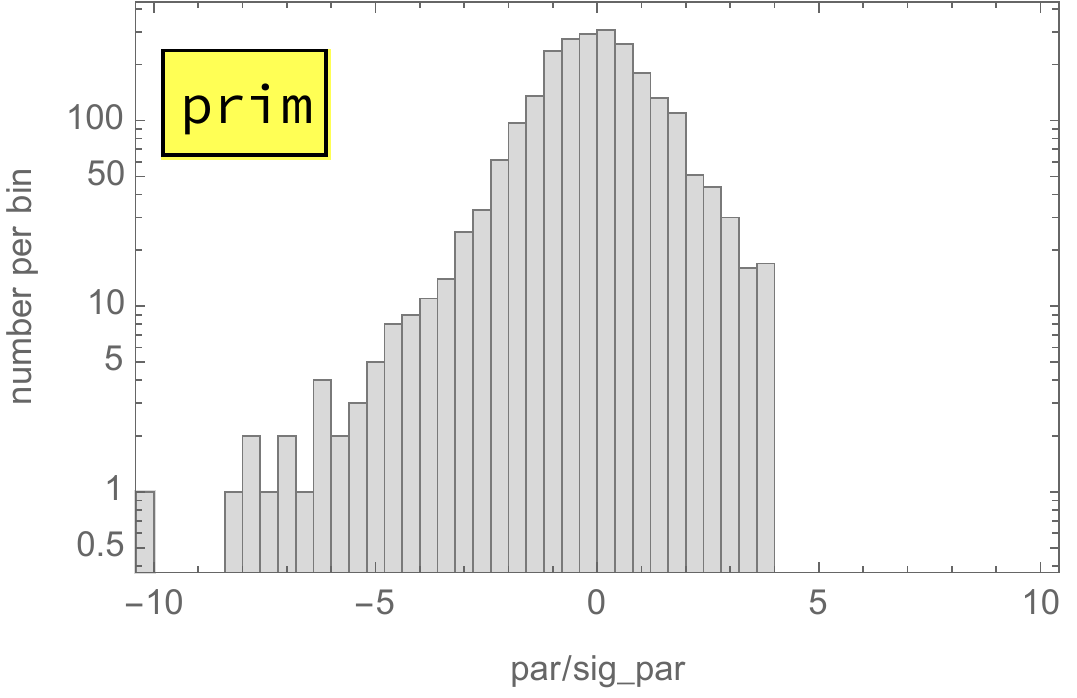}
    \includegraphics[width=0.47 \textwidth]{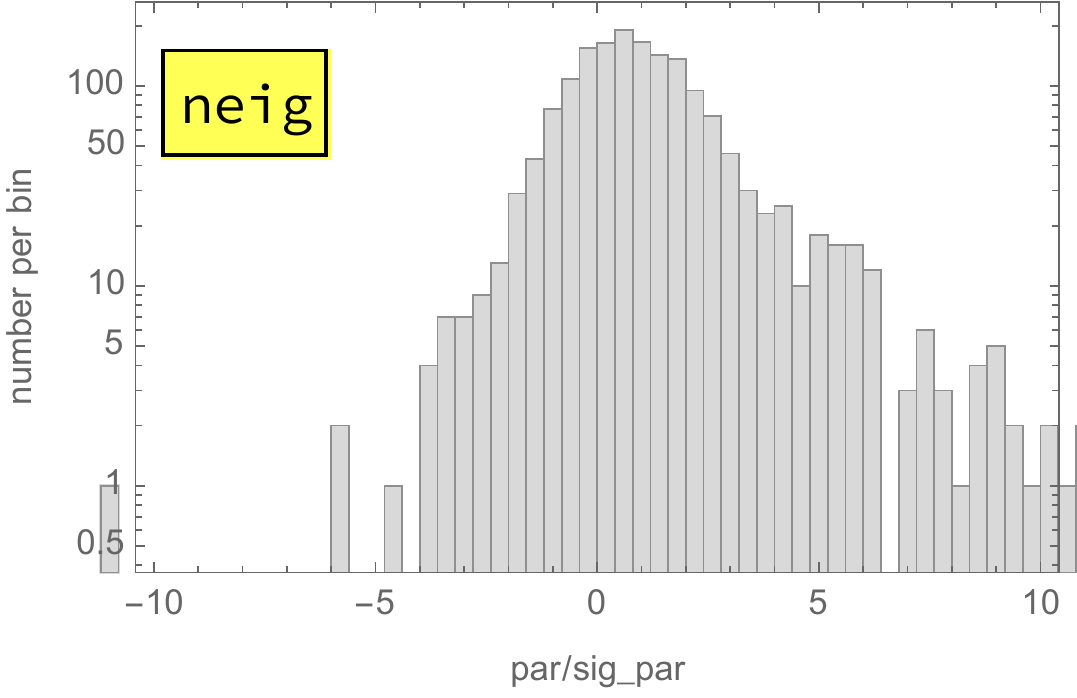}
    \caption{Distributions of normalized parallax $\varpi/\sigma_\varpi$ for primary MIRAGN sources selected to CDAGN and their nearest neighbors within $2\arcsec$ separation {right}.}
    \label{par.fig}
\end{figure*}

This selection procedure is expected to remove $1-2^2/(11^2-5^2)\simeq 96$\% of stellar contaminants randomly aligned with quasars within $2\arcsec$. However, if the rate of physical dual quasars within 2$\arcsec$ is much lower than the rate of chance stellar neighbors, the selected sample may still include many false positives. To estimate the rate of stellar contamination, the normalized parallaxes of the primaries (quasars) and all the nearest companions are investigated. The distributions are shown in Figure~\ref{par.fig}. The cut for the primaries is just one of our filters applied at an earlier stage to discard some of the stellar interlopers. Both distributions are sharply peaked at $\simeq 0$, although the neighbors' histogram seems to be slightly shifted to positive normalized parallax. We find a weak sign of a secondary bump at $\varpi/\sigma_\varpi\sim 5$ for the neighbors, which may be caused by the remaining stellar companions. The total number of neighbors with $\varpi/\sigma_\varpi\ >4$ is 144. Very approximately, the actual number of stellar contaminants may be twice this number if the parallaxes are perturbed symmetrically around the modal value. The estimated rate of stellar companions from the observed distributions of parallax is 12\%. We do not use the parallaxes of resolved neighbors to clean the final selection, however, because this astrometric parameter is often not available at close separations.

\begin{figure*}
    \includegraphics[width=0.47 \textwidth]{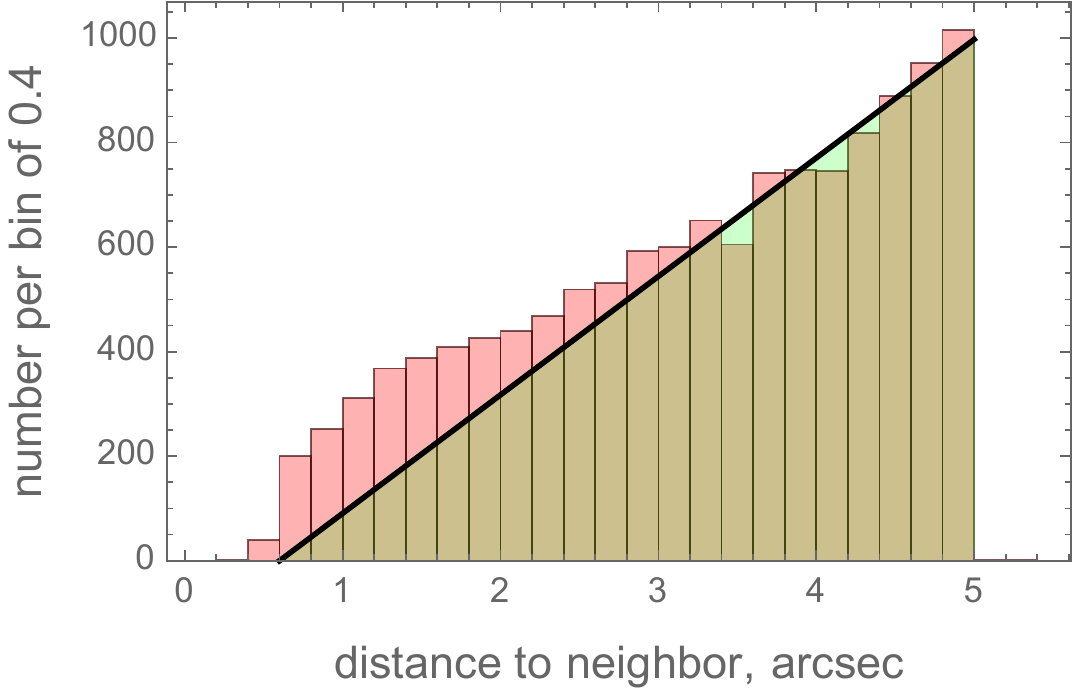}
    \caption{Histogram of neighbor distances for  $12,049$ MIRAGN--Gaia sources with at least one neighbor within $5\arcsec$ but no neighbors between $5\arcsec$ and $11\arcsec$. The straight line represents the statistically expected rate of chance neighbors taking into account the hard limit on the resolving capabilities of Gaia.}
    \label{neig5.fig}
\end{figure*}

As an additional and independent verification method, we employ near-neighbor statistics. The intermediate sample of $12,049$ MIRAGN--Gaia sources with at least one neighbor within $5\arcsec$ but no neighbors inside the ring between $5\arcsec$ and $11\arcsec$ is used. As previously discussed, this selection favors sources located in low-density areas, which greatly improves the reliability of statistical estimation. The sample sources have $12,713$ neighbors, i.e., most of them have only one neighbor. Fig. \ref{neig5.fig} displays the histogram of separations for this sample with a bin width of $0\farcs2$. The absence of neighbors closer than $\sim 0\farcs6$ is caused by the hard limit on angular separation in Gaia. The straight line originating at the limiting separation represents the rate of random neighbors assuming that the neighbors between $4\arcsec$--$5\arcsec$ are all random. We see a definite excess of neighbors at separations below $\sim 3\arcsec$. Counting the number of neighbors above the expected rate with separations $<2\arcsec$, the estimated fraction of statistical interlopers is 46\%, leaving the 54\% of the Sample 2 as physical dual sources. The estimate for the rate of stellar interlopers is much higher than the previously obtained 12\%, possibly because the near-neighbor estimation is overly conservative. It is known that the resolving capabilities of Gaia EDR3 are strongly degraded already at separations below 
$<2\arcsec$ \citep{2021A&A...649A...5F}. The actual rate of chance neighbors is a concave function rather than a straight line. Also, some of the neighbors at $>4\arcsec$ may still be genuine double AGNs because Sample 2 is not limited to high-redshift sources.

\begin{deluxetable*}{lcclr} \label{format.tab}
\tablehead{\colhead{column} & \colhead{parameter} & \colhead{unit} & \colhead{meaning} & \colhead{note} }
\startdata
1 & WISEA name &  &   & from \citet{2015ApJS..221...12S} \\
2 & RA & degrees & right ascension  & from Gaia EDR3\\
3 & DE & degrees & declination  & from Gaia EDR3\\
4 & source id &  &  & from Gaia EDR3 \\
5 & $G$ & mag & $G$ magnitude & from Gaia EDR3 \\
6 & $W1$ & mag & $W1$ magnitude & from WISE \\
7 & $W1-W2$ & mag & $W1-W2$ color & from WISE \\
8 & phe &  & \phe\ & from Gaia EDR3, empty if not available \\
9 & $z$ &  & redshift & from SDSS/BOSS survey, empty if not available \\
10 & $z_{\rm pre}$ &  & ML-predicted redshift & this paper, -1 if not available\\
11 & $\baz_{\rm cla}$ &  & ML-classified redshift & this paper, -1 if not available\\
12 & sep & arcsec & separation from nearest neighbor & this paper, -1 if greater than $2\arcsec$
\enddata
\caption{Format and layout of CDAGN catalog (published online).}
\end{deluxetable*}

\section{The merged CDAGN catalog}

The two output samples of candidate dual sources from the high-$z$ selection with elevated \phe\ values (Sample~1, 1023 sources) and objects with Gaia-resolved neighbors within $2\arcsec$ (Sample~2, 2363 sources) are merged to yield the final CDAGN catalog. At this final stage of catalog production, we identified 23 sources in Sample~1, which have $z_{\rm obs}-z_{\rm pre}>2$. Visual inspection of the images available through the online Pan-STARRS cutout service\footnote{\url https://ps1images.stsci.edu/cgi-bin/ps1cutouts} and SDSS spectra available through the online SDSS DR16 service \footnote{\url http://skyserver.sdss.org/dr16/en/tools/chart/navi.aspx} revealed that many of these cases are in fact gross errors in SDSS redshifts, as discussed in Section~\ref{gross.sec}. Furthermore, 219 objects were found to be present in both samples, and the duplicates were removed. 

At the final stage of filtering, we checked the CLASS metadata field, which contains the spectroscopic classification determination for those sources that are in the SDSS survey. Four sources emerged with a classifier STAR in that field. Source J081308.58+480643.2   at position $(123\fdg28584401, 48\fdg11200684)$ has a 16.8 mag companion resolved by Gaia at $1\farcs8$, which is indeed a fast-moving white dwarf. Several decades ago, it was much closer to the object of interest, because its proper motion is generally away from it.
The $1\farcs042$ companion of J095438.95+443356.3 is brighter than the primary target by about 1 mag and is also moving roughly away from it. Spectroscopically classified as a carbon star, the spectrum may be of composite nature with broad absorption features and a single powerful emission line. The source J154653.67+573533.9 is an interesting example of a legitimate quasar classified as BAL by \citet{2006ApJS..165....1T} whose spectrum can be contaminated by a likely stellar companion separated by $1\farcs511$. Some of the objects in the intriguing class of quasars with flat and featureless spectra may be such blends. The $1\farcs797$ companion to the quasar J223223.70+135434.6 is a high proper motion star ($42$ \masyr) of unknown type. After removal of the four identified stellar contaminants,
the final merged sample of 3140 sources constitutes the CDAGN catalog, which is published online.

Table \ref{format.tab} describes the format of CDAGN. Not all of the values are available in some of the columns. The \phe\ parameter is not available in Gaia EDR3 for 479 sources, which come from Sample~2 where \phe\ was not used in the selection process. Spectroscopic redshifts in column 9 are not available for 2814 sources. ML-predicted and ML-classified redshifts in columns 10 and 11 are not specified for 521 object originating from Sample~2. Separations in field 12 are not available for the entire Sample~1 contribution of 781 objects. The \{0.158655, 0.841345\} quantiles of $z_{\rm obs}-z_{\rm pre}$ for 301 sources with both values of redshift available, which correspond to $\pm \sigma$ for a Gaussian distribution, are \{-0.24, +0.51\}. The strong asymmetry is caused by the skewed sample distribution having a heavy tail toward positive differences.

\section{Discussion and summary}
The published CDAGN catalog includes 3140 objects. One quarter of it (781 entries) have $-1$ in column 12 for separation to the nearest companion, i.e., these objects come from Sample 1 and are not resolved by Gaia at small angular separations. The main risk associated with this part of the catalog is that some of Sample 1 objects are not dual or double at all, and the elevated level of \phe\ is caused by other circumstances. Indeed, a small number of Galactic planetary nebulae and Herbig Ae/Be stars have sneaked into our selection (Section \ref{stars.sec}) with very high values of \phe\ apparently triggered by the presence of bright emission structures surrounding the central stars. Our ML redshift determination produced incorrect results for such exotic objects because of the unusual photometric properties (large mid-IR excess coupled with blue optical colors). Several percent of this sample may also be nearby AGNs with extended image components, such as substrate galaxies, which propagated into the final catalog due to ``regular" errors of the ML redshift prediction and classification. A smaller fraction represents nearby AGNs with grossly incorrect spectroscopic redshifts (Section \ref{gross.sec}). The user is advised to pay attention to discrepant values of $z_{\rm obs}$ and $z_{\rm pre}$, which may indicate such problems. Given these risks and the relatively small representation of Sample 1, one may ask if its inclusion is warranted. Our motivation is that this part may capture candidate dual sources at separations beyond the current angular resolution capabilities of Gaia EDR3. The smallest separation recorded in the catalog is $0\farcs388$; however, the resolving capabilities of Gaia EDR3 are known to be severely degraded at separations below $1\arcsec$. We hope that this selection captures some dynamically evolved dual AGNs that have already converged to the gravitational potential well minimum, which is a barely known area at present. Checking extant and collecting new high-quality images of CDAGNs is also quite helpful in avoiding remaining contaminants of Sample 1 provenance because the extended structures surrounding AGNs at $z<0.5$ are often clearly visible.

A higher degree of reliability of CDAGNs coming from Sample 2 appears to be guaranteed by the fact the these sources have already been resolved in Gaia EDR3 with separations within $2\arcsec$. However, a double source is not necessarily a dual AGNs. The fraction of optical pairs, which must be, on statistical grounds, mostly chance alignments with foreground stars, is difficult to accurately estimate. We use two indirect methods of estimation, which produce somewhat different results. This fraction amounts to roughly 12\% from the apparent distribution of parallaxes for the near neighbors and 46\% from the general near-neighbor distance considerations. Very conservatively, we conclude that more than half of the resolved sources in Sample 2 are genuine dual, i.e., physical, AGNs. As for the overall rate of double sources (including remaining stellar companions), we note that 2359 sources out of 3140 come from Sample 2, that is, they are already resolved at separations $<2\arcsec$, while not more than $\sim 8\%$ of the remaining 781 objects originating from Sample 1 have redshifts smaller than 1, which may trigger a false positive. Therefore, not more than 2\% of the final catalog are expected to be single, relatively nearby sources.

\begin{acknowledgements}
This work has made use of data from the European Space Agency (ESA) mission
{\it Gaia} (\url{https://www.cosmos.esa.int/gaia}), processed by the {\it Gaia}
Data Processing and Analysis Consortium (DPAC,
\url{https://www.cosmos.esa.int/web/gaia/dpac/consortium}). Funding for the DPAC
has been provided by national institutions, in particular the institutions
participating in the {\it Gaia} Multilateral Agreement.

This research made use of Astropy,\footnote{http://www.astropy.org} a community-developed core Python package for Astronomy \citep{2013A&A...558A..33A, 2018AJ....156..123A}, and \textsc{topcat} \citep{2005ASPC..347...29T}

Funding for the Sloan Digital Sky Survey IV has been provided by the Alfred P. Sloan Foundation, the U.S. Department of Energy Office of Science, and the Participating Institutions. 

SDSS-IV acknowledges support and resources from the Center for High Performance Computing  at the University of Utah. The SDSS website is www.sdss.org.

SDSS-IV is managed by the Astrophysical Research Consortium for the Participating Institutions of the SDSS Collaboration including the Brazilian Participation Group, the Carnegie Institution for Science, Carnegie Mellon University, Center for Astrophysics | Harvard \& Smithsonian, the Chilean Participation Group, the French Participation Group, Instituto de Astrof\'isica de Canarias, The Johns Hopkins University, Kavli Institute for the Physics and Mathematics of the Universe (IPMU) / University of Tokyo, the Korean Participation Group, Lawrence Berkeley National Laboratory, Leibniz Institut f\"ur Astrophysik Potsdam (AIP),  Max-Planck-Institut f\"ur Astronomie (MPIA Heidelberg), Max-Planck-Institut f\"ur Astrophysik (MPA Garching), Max-Planck-Institut f\"ur Extraterrestrische Physik (MPE), National Astronomical Observatories of China,New Mexico State University, New York University, University of Notre Dame, Observat\'ario Nacional / MCTI, The Ohio State University, Pennsylvania State University, Shanghai Astronomical Observatory, United Kingdom Participation Group, Universidad Nacional Aut\'onoma de M\'exico, University of Arizona, University of Colorado Boulder, University of Oxford, University of Portsmouth, University of Utah, University of Virginia, University of Washington, University of Wisconsin, Vanderbilt University, and Yale University.

The Pan-STARRS1 Surveys (PS1) and the PS1 public science archive have been made possible through contributions by the Institute for Astronomy, the University of Hawaii, the Pan-STARRS Project Office, the Max-Planck Society and its participating institutes, the Max Planck Institute for Astronomy, Heidelberg and the Max Planck Institute for Extraterrestrial Physics, Garching, The Johns Hopkins University, Durham University, the University of Edinburgh, the Queen's University Belfast, the Harvard-Smithsonian Center for Astrophysics, the Las Cumbres Observatory Global Telescope Network Incorporated, the National Central University of Taiwan, the Space Telescope Science Institute, the National Aeronautics and Space Administration under Grant No. NNX08AR22G issued through the Planetary Science Division of the NASA Science Mission Directorate, the National Science Foundation Grant No. AST-1238877, the University of Maryland, Eotvos Lorand University (ELTE), the Los Alamos National Laboratory, and the Gordon and Betty Moore Foundation.

\end{acknowledgements}
 
\facilities{Gaia, WISE, Sloan, PS1}

\software{Astropy \citep{2013A&A...558A..33A, 2018AJ....156..123A}, \textsc{topcat} \citep{2005ASPC..347...29T}}

\bibliography{manuscript}
\bibliographystyle{aasjournal}

\appendix
\section{Objects of note and peculiar sources}

In the process of compiling CDAGN and verifying its contents, we have encountered objects of special interest and oddities of different kinds. The frequency of such cases is elevated in CDAGN because the filters applied tend to select most perturbed objects with unusual properties, which are otherwise quite rare in the original MIRAGN sample. In this Section, some of the cases are listed and briefly discussed, irrespective of whether they have been included in the final catalog or rejected.

\subsection{Stellar objects mimicking quasars}
\label{stars.sec}
A small fraction of Galactic stars is present in the initial MIRAGN/EDR3 sample. These are very rare objects with unusual photometric properties, namely, extremely red $W1-W2$ and $W2-W3$ colors. Some of them are also associated with marginally high \phe\ values, as we discovered in the process of visual inspection of the most conspicuous cases.
\begin{itemize}
    \item Source J200236.33+173650.6   at position $(300\fdg65141639, 17\fdg61410948)$ is the known planetary nebula (PN) G056.8$-$06.9
    also known as K 3-51 \citep{1992secg.book.....A}. Gaia EDR3 lists multiple sources around this object within $10\arcsec$, which are likely triggered by clumps in the nebula. The nebula has a ring-like shape with a looped filament and a nearby small structure, which may be a very red companion. The \phe\ value is exceptionally high at 2.73 possibly due to the optically bright extended image of the nebula. 
    \item Source J070006.62+121441.5 at position $(105\fdg02763308, 12\fdg24479321)$ is the PN G202.9+07.4 = KN 60 of remarkable bipolar structure and a nearby companion separated by $1\farcs332$ from the central star, recently identified in Gaia data \citep{2020A&A...638A.103C, 2021A&A...656A..51G}. The combination of an optically bright two-lobed nebula stretching over at least $7\arcsec$ and a relatively bright neighbor produced one of the highest values \phe\ $=4.99$ in the initial sample. This object was previously misidentified as an AGN in a number of papers. 
    \item Source J034732.98+350248.8 at position $(56\fdg88742456, 35\fdg04684135)$ is the well-known IC 351 \citep[e.g., ][]{2001A&A...378..843K} of ring-like elongated shape and a few clumps, which resulted in three bogus companions in Gaia EDR3 resolved within $3\arcsec$.  Like many other PNs, it is a rather bright radio source with a peak flux of $\sim 30$ mJy listed in the NVSS \citep{1998ApJS..117..361C}. 
    \item Source J184530.00-251214.6 at position $(281\fdg37502112, -25\fdg20409959)$ is an apparently new PN not listed in SIMBAD or any relevant catalogs, which has been misclassified as an AGN in a few compilations. It has a remarkable bipolar structure and an hour-glass shape stretching over at least $7\arcsec$. As in the previous cases, Gaia EDR3  resolves a few close neighbors, which are likely some clumps within the bright nebula. 
    \item Source J181933.56+035449.6 = IRAS 18170+0353 at position $(274\fdg88982401, 3.91378495)$ is an extreme infrared object and a know SiO and OH maser \citep{2010PASJ...62..525D, 2015A&A...582A..68E}. It has a stellar morphology without any resolved structure in available images. Gaia EDR3 presents a statistically significant proper motion of this object, while the measured parallax is statistically consistent with zero. 
    \item
    Source J060654.83+203916.2 at position $(91\fdg72849084, 20\fdg65453833)$ is not listed in SIMBAD but it has been incorrectly included in a number of AGN lists, as well as in the catalog of suspected YSOs by \citet{2016MNRAS.458.3479M}. It was recently classified as a new Herbig Ae/Be star by \citet{2020A&A...638A..21V} using a machine learning approach. Composite Pan-STARRS images suggest the presence of a shell-like structure around the central point source. This object seems to be an example of apparently isolated young star as there are no other known YSOs in the close vicinity.
    \item
    Source J185555.28+081701.7 at position $(283\fdg98035277, 8\fdg28384173)$ is another seemingly isolated Herbig Ae/Be object not listed in SIMBAD \citep{2020A&A...638A..21V}. Composite Pan-STARRS images reveal at least three faint companions, which may be bright nebulae or protostar cores.
    \item
    Source J004449.74+632251.0 = IRAS 00418+6306 at position $(11\fdg20729618, 63\fdg38083536)$ has been identified as a Herbig Ae/Be star \citet{2020A&A...638A..21V} and an OH maser \citep{2015A&A...582A..68E}. There is a hint of a blue shell around the star in composite images.
    \item
    Source J160028.60-562843.9 = IRAS 15564-5620 at position $(240\fdg11916884, -56\fdg47886695)$ has been identified as a Herbig Ae/Be star \citet{2020A&A...638A..21V} and an OH maser \citep{2015A&A...582A..68E}. It is listed in Gaia EDR3 with a small parallax $0.23$ mas but a surprisingly high proper motion $6.38$ \masyr. Its color is an outstanding \gbp$-$\grp$=4.00$ mag. The location in a relatively crowded part of the sky close to the Galactic plane and the presence of other very red objects with similar proper motions suggests a young association or a star-forming region.

\end{itemize}

\subsection{Quasars with erroneous SDSS spectroscopic redshifts}
\label{gross.sec}
This category of sources came under our consideration because of the procedures used in collecting the Sample~1. Initially based on ML-predicted and classified redshifts above 1, we also decided to back-substitute all objects from the SDSS sample with spectroscopic redshifts above 1. There may be several cases of greatly overestimated SDSS redshifts in CDAGN. A possible explanation is a blend of relatively nearby galaxies with AGNs with a well-aligned foreground star.

\begin{itemize}
    \item Source J003005.38+295708.0 at position $(7\fdg52241384, 29\fdg95221128)$ has a continuum spectrum peaked at approximatel 550 nm and a single broad and emission line at 755 nm, which was incorrectly interpreted as a Ly$\alpha$ line in the SDSS pipeline. This radio-loud AGN undoubtedly resides in a relatively nearby spherical galaxy, as revealed by available images. The spectroscopic SDSS-determined redshifts 5.2 is incorrect, while the ML-predicted redshift 0.15 is closer to the truth. This implies that the dominating emission feature is the H$\alpha$ line.
    \item
    Source J142813.79+311417.1 at position $(217\fdg05751109, 31\fdg23810507)$ is a similar case to the previous object. Its continuum is not normal for a quasar rising toward the red end and interspersed with narrow absorption and emission lines. The single broad emission line at 740 nm was misinterpreted as a Ly$\alpha$ line leading to a redshift value of 5.1. Pan-STARRS images show an extended host galaxy with a hint at some structure in the color distribution. The ML-predicted redshift is 0.50.
    \item Source J083333.66+234943.6 at position $(128\fdg390271593514, 23\fdg8287733729836)$ has a stellar-like spectrum with a bump in the blue part and a single very broad and complex emission line at 930 nm. Interpreted as a Ly$\alpha$ line, it leads to one of the highest SDSS-determined redshifts of 6.7. The ML-predicted redshift is 0.64.
\end{itemize}

\subsection{Known and new gravitational lenses}
The filtering procedures in selection of Sample-2 of resolved candidate dual AGNs (Section \ref{reso.sec}) selects objects with a few tightly packed images within a circle of $5\arcsec$ radius but no close neighbors outside this circle. This favors multiply imaged strong gravitational lenses, which are rare and important objects in their own right. It is not surprising that we find an elevated rate of lensed images upon a limited visual inspection of Sample-2 objects. The list below includes only some of the objects in this category with a short description as appropriate.

\begin{itemize}
    \item Source J111816.92+074558.5 = QSO 1115+080 at position $(169\fdg57066749, 7\fdg76625047)$ is a known lensed system of images \citep{2012ApJ...755...31C, 2017MNRAS.466.3088W, 2018A&A...618A..56D, 2019A&A...622A.165D} at $z=1.73$. Available HST NICMOS/NIC2 images show the complex configuration with a row of closely packed images on the East side and two wider and fainter images on the West side of the lens, which is barely discernible in Pan-STARRS images. Gaia resolved four of these images.
    \item
    Source J125149.16-122217.1 at position $(192\fdg954872187, -12\fdg37147381)$ is likely to be a previously undetected gravitational lens. Composite Pan-STARRS images reveal an extended image probably composed of at least three images with a complex color distribution. Gaia resolves only two sources separated by $0\farcs9$, however, and the northern companion has a large measured proper motion of 8.4 \masyr. The ML-predicted redshift is 1.11.
    \item Source J205144.29+252736.9 at position $(312\fdg93464628, 25\fdg46031899)$, with three images resolved within $5\arcsec$ in Gaia, has been selected as a candidate strong lens by \citet{2019A&A...622A.165D}. Pan-STARRS images show, however, that the two closely separated ($0\farcs703$) components have much different colors, so that the blue one is the quasar and the red one at position angle $204\degr$ may be an interloping star. The astrometric measurements for the red companion are quite uncertain. The quasar with $z_{\rm pre}=2.58$ is known to be optically variable \citep{2018JAD....24....3U}, so follow up observations on the light curves of the components can confirm or rule out this suggested lens.
    \item Source J054535.89+582805.8 at position $(86\fdg39963598, 58\fdg468300681)$ is likely a new gravitational lens. It has not been selected by previous Gaia-based search engines because only two close images have been resolved at separation $0\farcs817$---however, more seems to be present in the available images. The measured proper motion of the northern component is a high 9.06 \masyr\ but Gaia proper motions are known to be strongly perturbed beyond the error budget by close unresolved multiplicity \citep{2017ApJ...840L...1M}.
    \item Source J055305.35+091052.5 at position $(88\fdg27226393, 9\fdg18136749)$ is similar to the previous case in that only two images have been resolved in Gaia EDR3 at separation $1\farcs106$ but Pan-STARRS images suggest more components. It may be a new gravitational lens.
    \item Source J121547.10+293409.8 at position $(183\fdg94628602, 29\fdg56940657)$ has been selected by \citet{2012AJ....143..119I} as a candidate lensed quasar based on SDSS data. Gaia EDR3 resolved this object into two images at one of the closest separations $0\farcs469$. 
\end{itemize}

\end{document}